\documentclass[aps, pra, singlecolumn]{revtex4}
\usepackage{amsmath}
\usepackage{dsfont}
\usepackage{graphicx}
\usepackage{graphicx,amsmath,amssymb}

\newcommand{\kk}{\mathbf{k}}

\newcommand{\ii}{\mathrm{i}}

\begin{document}

\title{Theory of damping in magnetization dynamics, dispelling a myth and pointing a way forward}

\author{ D M Edwards}

\address{Department of Mathematics, Imperial College London, London SW7~2BZ, United Kingdom}


\begin{abstract}

There is a widely-held belief amongst theoreticians that the Gilbert damping parameter $\alpha$ in magnetization dynamics is infinite for a pure metal at T=0. The basic error leading to this belief is pointed out explicitly and the various methods of calculation used are viewed in a unified way based on the Lorentzian lineshape of ferromagnetic resonance spectra. A general torque formula for $\alpha$ is proposed as a good starting-point for treating inhomogeneous materials such as alloys, compounds and layered structures.
Local spin density functional theory provides a simple physical picture, in terms of a non-uniform precessional cone angle in ferromagnetic resonance, of how such inhomogeneity contributes to the damping. In a complementary many-body theory this contribution is given by a vertex correction to the torque-torque response function.

\end{abstract}

\maketitle

The damping of magnetization dynamics in ferromagnetic metals and alloys is of critical importance in spintronic devices. Damping largely controls the speed at which a device can operate and its energy requirement. In device physics damping is usually treated phenomenologically by means of a Gilbert term in the Landau-Lifshitz-Gilbert equation~\cite{LLG,Gil} and many quantum-mechanical calculations of the Gilbert parameter have been made for specific materials~\cite{GIS, LMCMB, LSYK, EMKP, Sak, SKBTB, MKWE, BCEU}. A reliable treatment of damping in transition metals and alloys would be an invaluable guide in the search for materials with very low damping~\cite{Schoen1, Lee}, as required for the future development of devices such as magnetic random access memory(MRAM). Most recent work in this direction is concerned with the important intrinsic contribution arising from spin-orbit coupling(SOC) and it is this which concerns us here. A satisfactory theory should work in the limit of a pure metal but almost all existing calculations predict that the Gilbert damping parameter $\alpha$ diverges to infinity for a pure metal at T=0. This would mean that in the pure metals Fe, Co and Ni at low temperature the linewidth in a ferromagnetic resonance (FMR) experiment would be much too large for the resonance to be observed. The prediction or acceptance of infinite damping has been made by so many authors~\cite{Kam, GIS, GM, LMCMB, LSYK, EMKP, Sak, SKBTB, MKWE, BCEU} over the last forty years that it has acquired the status of a myth. A very recent paper~\cite{G} repeats it once again. It is noteworthy that no experimentalist seems to have troubled to investigate the problem by work on high purity metals and dilute alloys at low temperature. The aim of this article is not only to dispel the myth but to formulate a firm starting-point for future calculations of $\alpha$ in technically important materials such as alloys, compounds and layered structures.

The most direct method to investigate damping,both experimentally and theoretically, is to study the ferromagnetic resonance (FMR) linewidth. In FMR a uniform static magnetic field $H$ is applied and the absorption of a transverse microwave field of angular frequency $\omega$ peaks around the frequency $b_{ex}/\hbar$ where $b_{ex}=2\mu_{B}H$ is the Zeeman energy. For $H$ in the z direction the absorption is determined by the imaginary part of the dynamical transverse susceptibility $\chi_{-+}(\omega)$. This susceptibility, which must include the effect of SOC, can be calculated by standard many-body theory using the Kubo formula or by time-dependent spin-density functional theory (SDFT). In practice the many-body method is usually based on a tight-binding approximation and employs the random phase approximation (RPA) with a short-range screened Coulomb interaction. This is then equivalent to a time-dependent Hartee-Fock mean-field theory. The long-range interaction can also be included if care is taken that it does not enter the exchange terms~\cite{Kim69, Kim73}. SDFT is approximated similarly as a time-dependent mean-field theory in the local spin density approximation (LSDA) and the long-range Coulomb interaction presents no problem since it is effectively screened in the exchange-correlation functional.  It is useful to consider both these methods in parallel. In a system with varying direction of magnetization SDFT is based on a density matrix of order 2~\cite{vBH} rather than just spin and particle densities. $\chi_{-+}(\omega)$ is then coupled to fifteen other response functions which determine the longitudinal spin susceptibility as well as the charge response and mixed charge-spin responses~\cite{WvB, Ed}. These last relate to phenomena like the spin-Hall effect. Some of these response functions, including the longitudinal spin susceptibility, involve the long-range Coulomb interaction importantly even in the absence of SOC~\cite{Kim69, Kim73, Ed}. Costa and Muniz~\cite{CM}, following an earlier paper~\cite{CMLKM}, show how SOC produces mode coupling in the RPA many-body approach. However the long-range Coulomb interaction is left untreated. Their paper is particularly important for being the first to challenge the myth of infinite damping.

We first discuss the case of a Bravais lattice which is appropriate for pure metals with a cubic structure like Fe and Ni at T=0. In both the approaches described above the dynamical susceptibility is related to mean-field susceptibilities of the general form
\begin{equation}\label{chi0}
\chi^{0}(\omega)
=N^{-1}\sum_{\kk mn}M_{mn}(\kk)
 \frac{f_{\kk n}-f_{\kk m}}{E_{\kk m}-E_{\kk n}-\hbar\omega+\ii\eta}.
\end{equation}
Here $E_{\kk m}$ is the energy of the one-electron state with wave-vector $\kk$ in band $m$, calculated in the presence of SOC, $f_{\kk m}$ is the corresponding occupation number, $M_{mn}(\kk)$ is a product of matrix elements and $\eta$ is a small positive constant which ultimately tends to zero. As in usual time-dependent perturbation theory equation~\eqref{chi0} represents the response to a perturbing field of angular frequency $\omega$ in which transitions occur between occupied and unoccupied states. "Intraband transitions" with $m=n$ clearly do not occur for $\omega\neq0$ owing to the cancellation of the Fermi functions. These transitions between identical states, which are not really transitions at all, can play no role in a dynamical process.
Hankiewicz et al have made a similar point~\cite{HVT}. However, in nearly all calculations of the Gilbert damping parameter $\alpha$, intraband transitions appear and lead to the infinite damping discussed above.

To dispel a myth effectively it is necessary to see how it has arisen. It is instructive to review, in a unified way, some methods which have been used to calculate $\alpha$ . We start from the Lorentzian form of the FMR lineshape which is well-established experimentally~\cite{Kal} and theoretically~\cite{CM}. Near the resonance the dynamical transverse susceptibility is dominated by a pole at $\hbar\omega=b_{ex}+\hbar\Delta\omega$ where  $\Delta\omega\sim\xi^{2}$,  $\xi$ being the SOC parameter, so that
\begin{equation}\label{chipole}
\chi_{-+}(\omega)=-\frac{2\langle S^{z}\rangle/N}{\hbar(\omega-\Delta\omega)-b_{ex}}.
\end{equation}
Here $S^{z}$ is the $z$ component of total spin and $N$ is the number of atoms in the crystal. Near the resonance the FMR absorption is determined by
\begin{equation}\label{Imchi}
\Im(\chi_{-+}(\omega))=-\frac{2(\langle S^{z}\rangle/N)\Im(\hbar\Delta\omega)}
{(\hbar\omega-\Re(\hbar\Delta\omega)-b_{ex})^{2}+(\Im(\hbar\Delta\omega))^2}.
\end{equation}
$\Re(\hbar\Delta\omega)$ corresponds to a shift in the resonance frequency and $\Im(\hbar\Delta\omega)$ determines the linewidth, both due to SOC. The Gilbert damping factor $\alpha$ is given by $\Im(\hbar\Delta\omega)/b_{ex}$ (e.g.~\cite{EW}). The most direct way to calculate $\alpha$ is a brute-force numerical RPA calculation of $\Im(\chi_{-+}(\omega))$, with SOC included, as a function of $\omega$ around the resonance. Costa and Muniz~\cite{CM} performed such calculations using the tight-binding approximation and found perfect  Lorentzians from which they deduced $\alpha$. Taking a monnolayer of Co as an example they found no tendency for $\alpha$ to diverge in the pure limit of sharp electronic states. This method of calculating $\alpha$ is very computer intensive and more economic methods exist if one assumes a Lorentzian curve from the outset.  

It follows immediately from~\eqref{chipole} that
\begin{equation}\label{alpha1}
\alpha=\Im(\hbar\Delta\omega)/b_{ex}=\frac{2\langle S^{z}\rangle}{Nb_{ex}}
\Im(\frac{1}{\chi_{-+}(b_{ex}/\hbar)}).
\end{equation}
This new formula for $\alpha$ may be regarded as exact. A full treatment of the transverse susceptibility includes coupling to other modes and leads to a rather complex expression in terms of sixteen mean-field susceptibilities of the form~\eqref{chi0} with different sets of matrix elements~\cite{Ed}. There is an enormous simplification in the case of a Bravais lattice if we calculate $\alpha$ only to second order in the SOC parameter $\xi$. Following the arguments of~\cite{Ed} it is readily found that coupling of the transverse susceptibility to other modes is then eliminated and that $\chi_{-+}$ in~\eqref{alpha1} may be replaced by the mean-field susceptibility $\chi_{-+}^{0}$. This elimination depends on inversion symmetry, which is a property of a Bravais lattice. Without this symmetry, coupling of the transverse susceptibility to other modes occurs in general even to order $\xi^{2}$, as discussed later. It follows further that to order $\xi^{2}$ 
\begin{equation}\label{alpha2}
\alpha=(N\Delta^{2}/2\langle S^{z}\rangle b_{ex})\Im(\chi_{-+}^{0}(b_{ex}/\hbar))
\end{equation}
where $\Delta$ is the exchange splitting in the band structure.
It is usually sufficient to calculate the last factor to first order in $b_{ex}$ so we may take the unphysical limit $b_{ex}\rightarrow 0$, but with due care as discussed below. Then
\begin{equation}\label{alpha3}
\alpha=(N\Delta^{2}/2\langle S^{z}\rangle)[\partial_{\omega}\Im(\chi_{-+}^{0}(\omega)]_{\omega=0}
\end{equation}
where the electronic state energies and matrix elements in $\Im(\chi_{-+}^{0}(\omega)$ are calculated with $b_{ex}=0$. Before proceeding to the static $\omega\rightarrow 0$ limit it is essential not to include contributions from "intraband transitions", as pointed out after~\eqref{chi0}. This precaution was not taken in~\cite{GM}, where a similar formula was obtained, so the spurious infinite damping for a pure metal appeared. Sometimes it is preferable to keep the physical non-zero Zeeman field to remove all danger of including intraband transitions. This also gives the option of calculating the frequency-swept FMR linewidth as a function of Zeeman field. This has been measured~\cite{Kal} and can be converted to a frequency dependence of $\alpha$. Such a dependence has been discussed by Costa and Muniz~\cite{CM}. However the low-field limit is usually sufficient and here we take the limit $b_{ex}\rightarrow 0$, with the precaution mentioned above, to compare with other theoretical work. Following ~\cite{GM}, but excluding intraband terms, we find the following two expressions for $\alpha$ at T=0:
\begin{equation}\label{alpha4}
\begin{split}
\alpha&=(\pi\Delta^{2}/2\langle S^{z}\rangle)\sum_{\kk}\sideset{}{'}\sum_{mn}
{\vert\langle\kk m \vert S^{-}\vert\kk n\rangle\vert}^{2}\delta(E_{\kk m}-E_{F})\delta(E_{\kk n}-E_{F})\\&=(\pi\xi^{2}/2\langle S^{z}\rangle)\sum_{\kk}\sideset{}{'}\sum_{mn}
{\vert\langle\kk m \vert T^{-}\vert\kk n\rangle\vert}^{2}\delta(E_{\kk m}-E_{F})\delta(E_{\kk n}-E_{F}).
\end{split}
\end{equation}
Here $\pmb{S}=(S^{x}, S^{y}, S^{z})$ is the total spin operator, $S^{-}=S^{x}-i S^{y}$, $\xi h_{so}$ is the total spin-orbit interaction, $T^{-}=[S^{-}, h_{so}]$ is a torque operator and $E_{F}$ is the Fermi energy. The prime on the sum over bands means $m\neq n$ and the sum over $\kk$ is to be carried out as an integral over the Brillouin zone as usual. As pointed out these expressions are only correct to order $\xi^{2}$ so that in the second expression we must evaluate the electronic states and energies with $\xi=0$. The prime on the summation sign may then be omitted since the $m=n$ terms are zero owing to inversion symmetry~\cite{BCEU}. The resulting expression, which can now be written in terms of one-particle Green functions if desired,  is just the version of Kambersky's torque formula~\cite{Kam} for a Bravais lattice derived in two ways by Edwards~\cite{Ed}. It is the mean-field approximation to a much more general formula~\cite{Ed}, valid for an ordered or disordered system,
 \begin{equation}\label{alpha5}
\alpha=-(\xi^{2}/2b_{ex}\langle S^{z}\rangle)\Im [\chi_{T}^{\xi=0}(b_{ex}/\hbar)].
\end{equation}
We shall refer to this as the general torque formula. It is exact to order $\xi^{2}$ and we have left open the option of taking the limit $b_{ex}\rightarrow 0$. Here the torque-torque response function is given by the Fourier transform of a retarded Green function using the Kubo formula
\begin{equation}\label{chiT}
\chi_{T}(\omega)=\int\langle\langle T^{-}(t),T^{+}\rangle\rangle e^{-\ii\omega t}dt. 
\end{equation}
The wide application of~\eqref{alpha5} is discussed later and we recall that the second expression in~\eqref{alpha4}, corresponding to the mean-field approximation to $\chi_{T}$, is only valid for a Bravais lattice. To evaluate the integral over $\kk$ in the formula~\eqref{alpha4} numerically it is usual to replace the delta-functions by Lorentzians of width proportional to an inverse relaxation time parameter $\tau^{-1}$. This broadening of the electron states may be regarded as a crude representation of the effect of impurity and/or phonon scattering. The limit $\tau^{-1}\rightarrow 0$ of a perfect crystal at T=0 leads to a finite value of $\alpha$ but is quite tricky to perform numerically~\cite{UE}. If we wrongly retain SOC in calculating the electron states in the second expression of~\eqref{alpha4} the diagonal matrix elements are non-zero and lead to the notorious infinite damping parameter $\alpha$. The only work which deals correctly with $\alpha$ in pure metals is reported in fig.1 of~\cite{BCEU} and in~\cite{CM,UE}.(In~\cite{UE} the caption of fig.1 should read "with and without SOC included in calculating electronic states").

We now turn to the task of establishing a firm basis for calculating the damping parameter $\alpha$ in technically important materials, which are typically random alloys or layered structures. This task is greatly simplified if we are satisfied with calculating $\alpha$ to second order in the SOC parameter $\xi$. This should be sufficient in nearly all systems of interest. At room temperature the $\xi^{2}$ dependence of $\alpha$ is well-established experimentally in several alloy systems, including some containing Pt with its large SOC~\cite{Scheck,He}. The general torque formula~\eqref{alpha5} is a very convenient starting-point. Its derivation in Appendix A of~\cite{Ed} is for a completely general ferromagnetic material, either ordered or disordered, and again relies only on the universal FMR Lorentzian lineshape. The derivation proceeds by comparing an exact relation between $\chi_{-+}(\omega)$ and $\chi_{T}(\omega)$ with an expansion of~\eqref{chipole} in the limit $\hbar\Delta\omega/(\hbar\omega-b_{ex}) \rightarrow 0$ followed by $\hbar\omega\rightarrow b_{ex}$. This order of limits is essential and results in the form~\eqref{alpha5} where $\chi_{T}$ is evaluated in the absence of SOC. A similar formula was derived by Kambersky~\cite{Kam} in another way where crucially the prescription $\xi=0$ did not become apparent. The formula is remarkable for describing the essence of a phenomenon arising solely from SOC without the need to include SOC in the calculation.

 The calculation of $\chi_{T}$ in a disordered system is still a very demanding problem. It may be approached using the RPA of standard many-body theory or, less obviously, using time-dependent LSDA. A diagrammatic RPA treatment of $\chi_{T}$ involves a sum of ladder diagrams and the first term, without an interaction line, corresponds to the mean-field approximation $\chi_{T}^{0}$. The remaining terms constitute a vertex correction and we have shown above that in a monatomic Bravais lattice this vanishes. In a disordered system like an alloy, or a metal at finite temperature in a frozen phonon picture, this is not the case. However this great simplification persists if, in a very crude approximation, the system is replaced, at the outset, by an effective medium with the full translational symmetry of the lattice but finite electron lifetime. We are then led to the Kambersky-like formula~\eqref{alpha4} for $\alpha$ with a Lorentzian broadening of the delta-functions determined by relaxation times which may be  dependent on spin and temperature. This is the background to a recent calculation of $\alpha$ in bulk Ni at room temperature~\cite{UE} which is in reasonable agreement with experiment. A proper treatment of $\chi_{T}$ in a disordered material must deal simultaneously with the RPA vertex correction and any vertex corrections which arise in connection with methods of taking a configurational average, such as the coherent potential approximation (CPA). There is a small literature on this problem as applied to $\chi_{-+}$, not $\chi_{T}$, in a one-band model~\cite{YS,SC}. Santos and Costa~\cite{SC} find that for dilute non-magnetic impurities the RPA vertex correction is particularly important. However as yet the many-body approach is far from being able to provide reliable results for $\alpha$ in real disordered materials. The time-dependent LSDA method seems more promising. As shown below, it gives a clear physical picture of the RPA vertex correction and separates it from the configurational averaging problem.

In a FMR experiment the local magnetization vector sweeps out a cone as it precesses around the Zeeman field direction and in the presence of SOC the cone angle $\theta(\pmb r)$ is a function of position. In the time-dependent LSDA $\theta(\pmb r)$ satisfies an integral equation whose solution is avoided in~\cite{GM} by taking a spatially-independent averaged cone angle $\overline{\theta(\pmb r)}$. This approximation enforces a uniform precession, as occurs in the absence of SOC, and removes the possibility of coupling between transverse and longitudinal susceptibilities.  It is very reasonable for a monatomic Bravais lattice where the variation of $\theta(\pmb r)$ within a unit cell is largely an artificial consequence of the local approximation. In the tight-binding framework of~\cite{Ed} it would not be an approximation at all for a monatomic Bravais lattice. However, in compounds, alloys and layered structures, variation of the cone angle between different types of atom and different layers may be very important. In the many-body approach the vertex correction in $\chi_{T}$ is the difference between the full $\chi_{T}$ and the mean-field approximation $\chi_{T}^{0}$. Since we have seen that the mean-field approximation works well for a homogeneous system, like a monatomic Bravais lattice, we conclude that the vertex correction corresponds to the effect of the spatial variation of the cone angle, which can be studied with the LSDA approach. This will be demonstrated explicitly in a forthcoming publication. This productive interplay between standard many-body theory and density-functional theory is quite unusual.

I would like to acknowledge a useful exchange of e-mails with Filipe Guimarães on the subject-matter of this paper.

\section*{}

\section*{}

\end{document}